\documentclass[10pt, conference]{IEEEtran}
\IEEEoverridecommandlockouts
\usepackage{cite}
\usepackage{amsmath,amssymb,amsfonts}
\usepackage{algorithmic}
\usepackage{graphicx}
\usepackage{textcomp}
\usepackage{xcolor}
\usepackage{verbatim}
\usepackage{balance}
\usepackage{booktabs}
\usepackage{multirow}
\usepackage{listings}

\lstdefinestyle{lists}{
    captionpos=b,
    abovecaptionskip=5pt,
    breaklines, 
    frame=single, 
    basicstyle=\ttfamily\tiny 
}
\graphicspath{ {./figures/} }

\def\BibTeX{{\rm B\kern-.05em{\sc i\kern-.025em b}\kern-.08em
    T\kern-.1667em\lower.7ex\hbox{E}\kern-.125emX}}

\begin{document}

\author{
    \IEEEauthorblockN{Seif Abukhalaf\IEEEauthorrefmark{1}, Mohammad Hamdaqa\IEEEauthorrefmark{1}, Foutse Khomh\IEEEauthorrefmark{2}}
    \IEEEauthorblockA{\IEEEauthorrefmark{1}Software and Emerging Technologies Lab (SAET), \IEEEauthorrefmark{2} SoftWare Analytics and Technologies Lab (SWAT)}
    \IEEEauthorblockA{Department of Computer and Software Engineering}
    \IEEEauthorblockA{Polytechnique Montréal, Montréal, Canada}
    \{seif.abukhalaf, mhamdaqa, foutse.khomh\}@polymtl.ca
}

\title{On Codex Prompt Engineering for OCL Generation: An Empirical Study}

\maketitle

\begin{abstract}
The Object Constraint Language (OCL) is a declarative language that adds constraints and object query expressions to Meta-Object Facility (MOF) models. OCL can provide precision and conciseness to UML models. Nevertheless, the unfamiliar syntax of OCL has hindered its adoption by software practitioners. 
LLMs, such as GPT-3, have made significant progress in many NLP tasks, such as text generation and semantic parsing. Similarly, researchers have improved on the downstream tasks by fine-tuning LLMs for the target task. Codex, a GPT-3 descendant by OpenAI, has been fine-tuned on publicly available code from GitHub and has proven the ability to generate code in many programming languages, powering the AI-pair programmer Copilot. One way to take advantage of Codex is to engineer prompts for the target downstream task. In this paper, we investigate the reliability of the OCL constraints generated by Codex from natural language specifications. To achieve this, we compiled a dataset of 15 UML models and 168 specifications from various educational resources. We manually crafted a prompt template with slots to populate with the UML information and the target task in the prefix format to complete the template with the generated OCL constraint. We used both zero- and few-shot learning methods in the experiments. The evaluation is reported by measuring the syntactic validity and the execution accuracy metrics of the generated OCL constraints. Moreover, to get insight into how close or natural the generated OCL constraints are compared to human-written ones, we measured the cosine similarity between the sentence embedding of the correctly generated and human-written OCL constraints. Our findings suggest that by enriching the prompts with the UML information of the models and enabling few-shot learning, the reliability of the generated OCL constraints increases. Furthermore, the results reveal a close similarity based on sentence embedding between the generated OCL constraints and the human-written ones in the ground truth, implying a level of clarity and understandability in the generated OCL constraints by Codex.
\end{abstract}


\section{Introduction}

Recent advances in deep learning and natural language processing have led to the development of large language models. These models have demonstrated improved performance in various downstream tasks, including code generation \cite{chowdhery2022palm, wang2021codet5}. Codex, developed by OpenAI, is a pre-trained large language model that has been fine-tuned on a vast corpus of publicly available code from GitHub repositories \cite{chen2021evaluating}. Prompt engineering techniques \cite{pre-train-prompt} are widely utilized to influence the behavior of large language models to perform downstream tasks.

Model-driven development (MDD) is a software development paradigm that uses models to increase the level of abstraction and guide the software development process. In MDD, developers use models to specify, design, and build software systems \cite{da2015model}. The Unified Modeling Language (UML) \cite{UMLspecs} is often used as a standard modeling language to facilitate communication and collaboration between developers and domain experts. In object-oriented design, UML is used to specify and visualize the classes and their associations. The Object Constraint Language (OCL) is a declarative language used to define clear and precise rules derived from the specifications as constraints on the UML models. This helps to improve the quality and efficiency of the model-driven software development process \cite{En2OCL}

When translating natural language specifications into OCL expressions, there can be multiple implementations. However, this can be a challenging and time-consuming task for modelers who are unfamiliar with the OCL syntax, especially in complex systems \cite{Cabot2006AmbiguityII, Wahler2008UsingPT, bajwa}. Similar to AI-assisted code generation tasks for different programming languages \cite{poesia2022synchromesh, BASHEXPLAINER}, OCL can also benefit from using large language models trained on code to generate constraints from natural language specifications. This can help modelers implement, validate, and review OCL constraints, making OCL language more accessible \cite{call_comm, bajwa_article}. The empirical study addresses the following research questions:
\newline

\begin{itemize}
    \item \textbf{RQ1. What is the impact of UML-enriched prompts on the generated OCL constraints?}
\end{itemize}

We address this question by examining the impact of different prompt designs on the reliability of OCL constraints generated from natural language specifications. We achieve this by using manually-crafted prompts in prefix shape and employing both zero- and few-shot learning methods \cite{pre-train-prompt}. The first design is a basic template without providing model information, while the rest are enriched with model classes and associations in both zero- and few-shot settings. 

To assess the generated OCL constraints, we first compile them using the UML-based Specification Environment (USE) modeling tool \cite{RichtersUSEtool} and report their syntactic validity. For the valid OCL constraints, we manually analyze their semantic correctness by reporting their execution accuracy. The analysis is conducted by two students, a master's and a Ph.D. student.
\newline

\begin{itemize}
    \item \textbf{RQ2. How similar are the generated OCL constraints by Codex to the ones written by humans?}
\end{itemize}

In this question, we focus on measuring the similarity between the correct OCL constraints written by humans and those automatically generated from the same natural language specifications. This could provide us with insights into how close or natural the generated OCL constraints are compared to human-written ones. To achieve this, we compute the sentence embeddings of both human and generated OCL constraints using a Transformer-based pre-trained language model and measure their cosine similarity.

The rest of this research paper is organized as follows: Section II provides the background and necessary terms. Section III explains our methodology. Section IV presents our empirical evaluation. Section V discusses our findings. Section VI introduces related work to the study. Section VII outlines the threats, and Section VIII concludes the paper.


\section{Background and Terminology}

This section provides background information on the concepts that are introduced in the empirical study.

\begin{figure}
    \centering
    \includegraphics[width=\columnwidth]{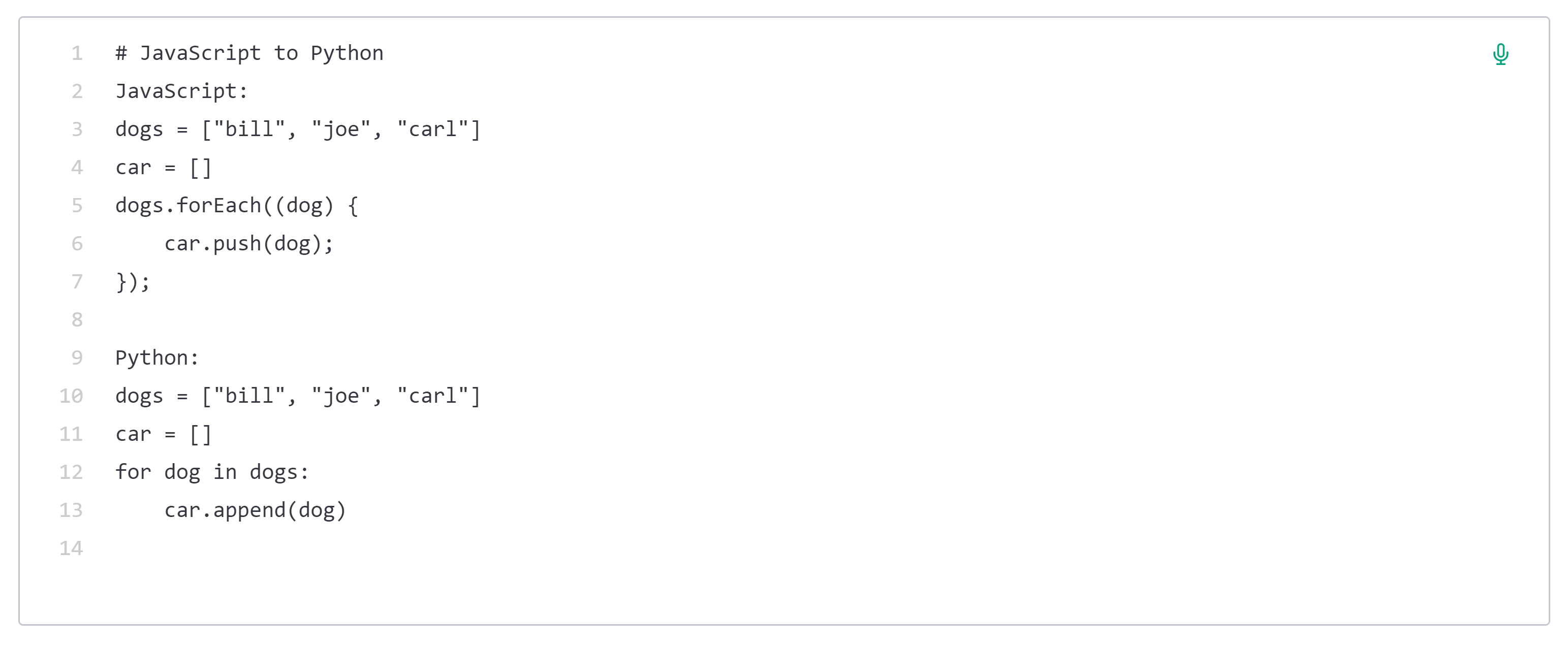}
    \caption{JavaScript to Python code transformation example.}
    \label{figure:playground}
\end{figure}

\textbf{Large Language Models (LLMs)} are deep neural networks that are pre-trained on massive amounts of text data to perform a wide variety of natural language processing (NLP) tasks, such as sentiment analysis, machine translation, and text generation \cite{pre-train-prompt}. One of the most recent advancements has been the development of Transformer-based models, such as GPT-3 (Generative Pre-trained Transformer 3) \cite{brown2020language-gpt}. Codex, a descendant of GPT-3, is fine-tuned on publicly available, massive code repositories from GitHub \cite{chen2021evaluating}. These repositories include code from multiple programming languages, such as Python, Java, and JavaScript, which improves the performance of Codex in code generation tasks for several programming languages. Codex powers GitHub Copilot, an AI-pair programmer designed to assist software developers \cite{nguyen2022empirical}. One way to leverage Codex is through the provided API or the playground by OpenAI. Figure \ref{figure:playground} shows an example of prompting Codex to perform a code transformation task from JavaScript to Python on the playground \footnote{https://platform.openai.com/playground}.

\textbf{Prompt Engineering} is a systematic approach to creating prompt functions that effectively communicate the target downstream task to pre-trained language models. The design and implementation of prompts have a crucial impact on the performance of the pre-trained language model \cite{pre-train-prompt}.

In the context of code generation, the prompt needs to be customized for the target programming language and clearly describe the task without ambiguity, as demonstrated in Codex \cite{chen2021evaluating}. Crafting prompts is an iterative refinement process that enhances the quality of generated code over time. The decisions to design prompts depend on the target task and the selection of the pre-trained language model. For example, prompts with the prefix shape are suitable for text generation tasks, while cloze-shaped prompts are more effective for sentiment analysis tasks. 

Another design consideration is to tailor the prompt template, which can be accomplished manually by human experts or through automatic approaches such as prompt mining to find the optimal prompt for the task. The template should be formulated in a way that aligns with the target task and the selected model. Prompting methods can also be used to explicitly train the model by enabling few-shot learning \cite{pre-train-prompt}, where a few examples, such as code snippets and test cases, are included in the prompt to influence the model regarding the target downstream task.


\textbf{Object Constraint Language (OCL)} is a declarative language used to specify rules for the structure and behavior of MOF models, including UML \cite{OCLspecs}. OCL is commonly used to express constraints such as class invariants, pre-conditions, and post-conditions. In the context of model-driven development (MDD), OCL can ensure formalism and consistency during software development \cite{CabotOCLGuide}.

\begin{figure}[t]
    \centering
    \includegraphics[width=\columnwidth]{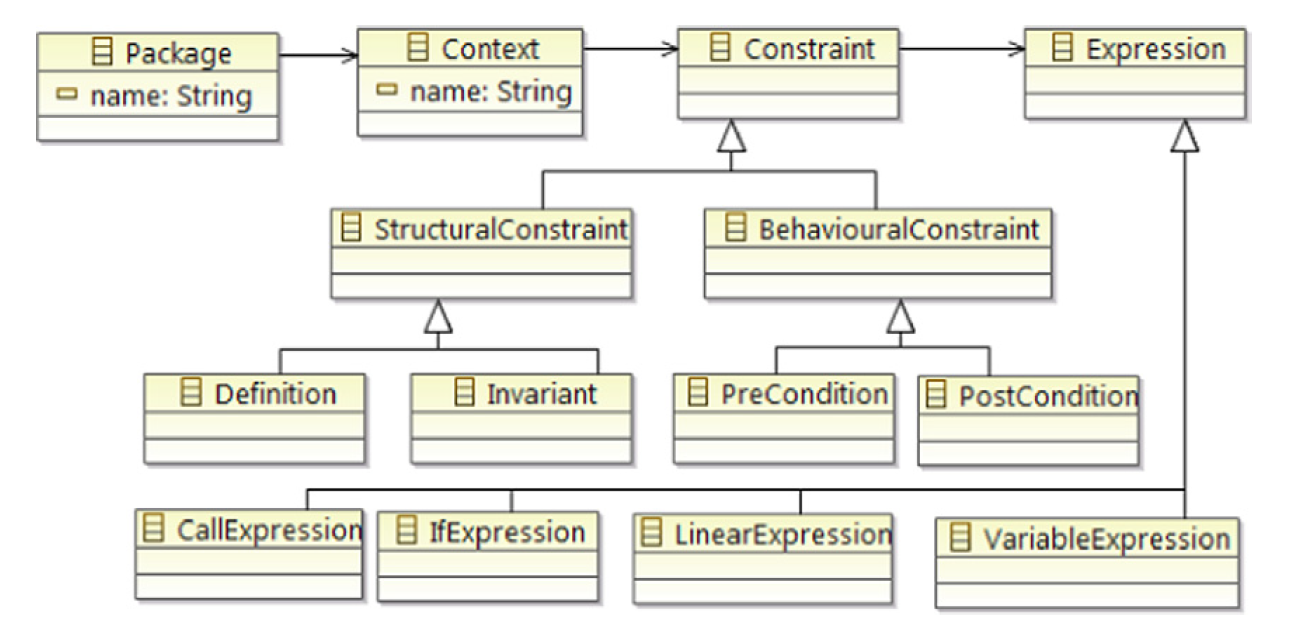}
    \caption{Overview of the OCL meta-model.}
    \label{figure:ocl-metamodel}
\end{figure}

The OCL meta-model defines the abstract syntax and semantics of OCL expressions. It organizes the basic elements of expressions in a hierarchical structure that allows for variations and extensions to the language, as shown in Figure \ref{figure:ocl-metamodel} \cite{bajwa_article}. By formalizing constraints using the OCL meta-model, tools can automatically check whether a model conforms to specified constraints and provide feedback to the user on violations or errors. The OCL abstract syntax defines the grammar and structure of the language. It is further divided into OCL types and OCL expressions. Common OCL types include data types, collection types, and message types. OCL expressions can include property expressions, if expressions, iterator expressions, variable expressions, and more.

\section{Methodology}

In this section, we outline the structure of our experiments that aim to analyze the OCL constraints generated by Codex. We begin by providing information on how our dataset was collected, followed by statistical analysis. Next, we present the three design considerations that were used in the prompts. Finally, we discuss the methods that were used to study the generated OCL constraints. 


\subsection{Data Collection \& Analysis}
Our primary motivation for collecting our dataset was the absence of suitable datasets \cite{call_comm} for assessing Codex in generating OCL constraints. Many existing datasets, such as the one published by Cabot on GitHub\footnote{https://github.com/jcabot/ocl-repository} and others in the literature \cite{mengerenki_1, mengerink2019empowering}, provide OCL constraints, but either contain few specifications in natural language or lack them entirely. To address this gap, we compiled a dataset of UML models and OCL constraints with their natural language specifications. This step enabled us to conduct experiments with different design settings and study the OCL constraints generated by Codex.

To obtain a dataset for our analysis, we conducted manual searches across various resources, including educational courses on model-driven engineering, literature \cite{Kleppe_Warmer, bc4j, bajwathesis}, and the GitHub repository \footnote{https://github.com/logicalhacking/ocl-examples}. This process resulted in 15 diverse UML models that are now ready for analysis. Table \ref{table:dataset-models-statistics} presents statistics on the number of UML classes, associations, and OCL specifications for each model in the dataset. In total, the dataset contains 94 classes, 101 associations, and 168 specifications written in natural language. The dataset has been made available on Zenodo to facilitate the study replication and provide resources for other researchers interested in the topic \footnote{https://doi.org/10.5281/zenodo.7749795}.

\begin{table}
\caption{An overview of the models in the dataset.}
\label{table:dataset-models-statistics}
\centering
\resizebox{\columnwidth}{!}{%
\begin{tabular}{@{}|llll|@{}}
\toprule
\multicolumn{1}{|c|}{\textit{\textbf{Model}}}    & \multicolumn{1}{c|}{\textit{\textbf{Classes}}} & \multicolumn{1}{c|}{\textit{\textbf{Associations}}} & \multicolumn{1}{c|}{\textit{\textbf{Specifications}}} \\ 
\bottomrule
\midrule
\multicolumn{1}{|l|}{\textit{Invoices}}          & \multicolumn{1}{l|}{2}                         & \multicolumn{1}{l|}{1}                              & 10                                                    \\ \midrule
\multicolumn{1}{|l|}{\textit{Train}}                      & \multicolumn{1}{l|}{2}                         & \multicolumn{1}{l|}{2}                              & 5
 \\ \midrule
\multicolumn{1}{|l|}{\textit{Mortgage}}                   & \multicolumn{1}{l|}{3}                         & \multicolumn{1}{l|}{3}                              & 7
\\ \midrule
\multicolumn{1}{|l|}{\textit{Tournament}}        & \multicolumn{1}{l|}{3}                         & \multicolumn{1}{l|}{3}                              & 9                                                     \\ \midrule
\multicolumn{1}{|l|}{\textit{Airport}}           & \multicolumn{1}{l|}{4}                         & \multicolumn{1}{l|}{5}                              & 7                                                     \\ \midrule
\multicolumn{1}{|l|}{\textit{Vehicle}}                    & \multicolumn{1}{l|}{5}                         & \multicolumn{1}{l|}{3}                              & 9                                            
\\ \midrule
\multicolumn{1}{|l|}{\textit{Person}}                     & \multicolumn{1}{l|}{6}                         & \multicolumn{1}{l|}{5}                              & 14                                          
\\ \midrule
\multicolumn{1}{|l|}{\textit{Health Records}}             & \multicolumn{1}{l|}{6}                         & \multicolumn{1}{l|}{6}                              & 5                                           
\\ \midrule
\multicolumn{1}{|l|}{\textit{Employment Agency}} & \multicolumn{1}{l|}{6}                         & \multicolumn{1}{l|}{6}                              & 26                                                    \\ \midrule
\multicolumn{1}{|l|}{\textit{Library}}                    & \multicolumn{1}{l|}{7}                         & \multicolumn{1}{l|}{6}                              & 6                                           
\\ \midrule
\multicolumn{1}{|l|}{\textit{Car Rental}}        & \multicolumn{1}{l|}{8}                         & \multicolumn{1}{l|}{8}                              & 5                                                     \\ \midrule
\multicolumn{1}{|l|}{\textit{Internet Service Provider}}  & \multicolumn{1}{l|}{9}                         & \multicolumn{1}{l|}{11}                             & 6       
\\ \midrule
\multicolumn{1}{|l|}{\textit{Royal \& Loyal}}    & \multicolumn{1}{l|}{9}                         & \multicolumn{1}{l|}{13}                             & 26                                                    \\ \midrule
\multicolumn{1}{|l|}{\textit{Business Relations}}         & \multicolumn{1}{l|}{11}                        & \multicolumn{1}{l|}{12}                             & 30                                                    \\ \midrule
\multicolumn{1}{|l|}{\textit{QUDV}}              & \multicolumn{1}{l|}{13}                        & \multicolumn{1}{l|}{17}                             & 3                                                     \\ 
\bottomrule
\midrule
\multicolumn{1}{|l|}{\textit{\textbf{Total}}}    & \multicolumn{1}{l|}{\textbf{94}}               & \multicolumn{1}{l|}{\textbf{101}}                    & \textbf{168}                                           \\ \bottomrule
\end{tabular}%
}
\end{table}


After conducting a thorough analysis of the OCL constraints in the models, we discovered a variety of OCL constraints. Specifically, there were 96 invariants, 15 pre-conditions, and 10 post-conditions. We also observed an imbalance in the distribution of invariants expressions compared to pre-and post-conditions, as shown in Table \ref{table:ocl-statistics-groups}. Among all the natural language specifications, only 54 of them did not contain their OCL constraints as ground truth. Therefore, we used the 168 specifications to generate OCL constraints in RQ1 and only used the 114 specifications with their ground truth to answer RQ2.

\begin{table}[b]
\centering
\caption{The frequency of OCL constraints in the dataset.}
\label{table:ocl-statistics-groups}
\resizebox{.75\columnwidth}{!}{%
\begin{tabular}{@{}|l|l|l|l|@{}}
\toprule
\textit{\textbf{Invariants}} & \textit{\textbf{Pre-conditions}} & \textit{\textbf{Post-conditions}} \\ 
\midrule
 96         & 15             & 10                 \\ \bottomrule
\end{tabular}%
}
\end{table}


\subsection{Prompt Design}
The approach we used to craft prompts was inspired by the systematic review conducted by Liu et al. \cite{pre-train-prompt}. To formulate the prompts, we considered the downstream task of code generation and took into account the following design considerations:

\textbf{Model Selection:} For text completion tasks such as code generation, auto-regressive large language models like GPT-3 \cite{brown2020language-gpt} can be utilized. Codex, which is a descendant of GPT-3 and has been fine-tuned on code repositories \cite{chen2021evaluating}, is an ideal candidate for generating OCL constraints based on natural language specifications. We selected Codex because it is accessible through the OpenAI API.

\textbf{Prompt Shaping:} The selection of Codex as the model and the downstream task influenced the shape of the prompts. Because the information sequence in the prompt comes entirely before the response, we chose the prefix shape for designing the prompts \cite{pre-train-prompt}. The prompts include the task description: “The task is to generate an 'object constraint language' (OCL) expression according to the given specification,” and the natural language specifications. In two of the experiments, the prompts are enriched with UML information about the model. All prompts end with the prefix “OCL:” to explicitly request Codex to complete the target OCL constraint.

\textbf{Prompting Approach:} We manually created a prompt template with a simple structure to gain insights into the OCL constraints generated by Codex in general. The template has placeholders to fill with the UML information of each model, the task description, and the natural language specification. OCL development tools, such as Eclipse, use the double dash “--” character to comment on OCL expressions. However, in the PlantUML format, the double dash is also used to describe the association between classes. To avoid conflicts, we replaced the double dash with the double slash “//” character and used it as the stop and restart sequence parameter for Codex to limit the generation to one expression.
    
\textbf{Training Strategy:} We mainly used two prompt designs without providing examples, i.e., zero-shot learning, and in the last prompt design, we enabled few-shot learning.


Given the design considerations, we performed three experiments with the prompt outlined below:

\subsubsection{\textbf{Basic Prompts}}
In our initial design, we crafted a straightforward prompt that included the task description, natural language specification, and the prefix “OCL:”. However, we chose not to include any relevant prior information regarding the UML models or provide any examples as few-shots. The experiment utilized a zero-shot learning approach \cite{pre-train-prompt}. This simple prompt enabled us to evaluate the OCL constraints generated by Codex without access to the UML information of the models, given its training on the publicly available GitHub repositories \cite{chen2021evaluating}. Listing \ref{listing:method-basic-prompt} is an example of the basic prompt we employed in the experiment. The first line specifies the task to request generating an OCL constraint, and the second line contains the natural language specification.

\begin{lstlisting}[caption={An example of a basic prompt with the task description, specification, and prefix.}, label={listing:method-basic-prompt}, captionpos=b, float=b]
// The task is to generate an object-constraint language (OCL) expression according to the given specification.

// Every customer who enters a loyalty program must be of legal age OCL:
\end{lstlisting}


\subsubsection{\textbf{UML Information \& Zero-Shot Learning}}
To improve the evaluation of OCL constraints generated by Codex, we created prompts with more information by using the Plant-UML syntax \cite{plantuml} to specify UML classes and associations, which we included in the prompt. To clarify the context, we added a sentence at the top of each prompt that references the relevant UML model. Listing \ref{listing:prompt-uml-zero} shows an example of this prompt design for the Royal \& Loyal case study. We did not provide any examples for this prompt design, meaning it will be a zero-shot learning setting.

\begin{lstlisting}[basicstyle=\tiny,caption={Illustration of the second prompt design, which incorporates UML specifications in the form of PlantUML.}, label={listing:prompt-uml-zero}, captionpos=b, float=t]
// This file containts the UML classes and associations of the model 'Royal & Loyal'.

// UML classes:
// class Loyalty Program{name: String, enroll(), getServices(): Service}
// class ProgramPartner {numberOfCustomers: Integer, name: String}
// class Membership{}
// class Customer{name: String, title: String, isMale: Boolean, dateOfBirth: Integer, age: Integer, cAge()} 
// class ServiceLevel{name: String}
// class Service{condition: Boolean, points Earned: Integer, points Burned: Integer, description: String, serviceNr: Integer}
// class CustomerCard{valid: Boolean, validFrom: Date, goodThru: Date, color: Color, printedName: String}
// class LoyaltyAccount{points: Integer, numbers: Integer, earn(), burn(), isEmpty()}
// class Transaction{date: String, amount: Integer, point: Integer, programs()}

// UML associations:
// association LoyaltyProgram 'programs 0..*' -- 'participants 0..*' Customer
// association LoyaltyProgram 'programs 1..*' -- 'partners 1..* ProgramPartner 
// association Loyalty Program 'programs 1' -- 'levels 1..*' ServiceLevel
// association Loyalty Program 'programs 0..1' -- 'membership 0..1' Membership
// association Customer 'owner 1' -- 'cards 0..*' CustomerCard
// association ProgramPartner 'partner 0..1' -- 'deliveredServices 0..*' Service
// association ServiceLevel 'currentLevel 1' -- 'current Level 0..*' Membership
// association ServiceLevel 'level 0..1' -- 'availableServices 0..*' Service 
// association Membership --> 'card 1' CustomerCard
// association Membership --> 'account 0..1' LoyaltyAccount
// association CustomerCard 'card 1' -- 'transactions 0..*' Transaction
// association LoyaltyAccount 'account 1' -- 'transactions 0..*' Transaction 
// association Service 'generatedBy 0..1' -- 'transactions 0..*' Transaction

// The task is to generate an 'object-constraint language' (OCL) expression according to the given specification. 
// Every customer who enters a loyalty program must be of legal age.
OCL:
\end{lstlisting}


\subsubsection{\textbf{UML Information \& Few-Shot Learning}}
In our last design, we utilized a few-shot learning approach, incorporating natural language specifications and OCL constraints as examples in the prompt. This approach was implemented automatically in our experiment. Few-shot learning has the potential to enhance the reliability of the code generated by Codex \cite{pre-train-prompt}.

\subsection{Analysis Methods}
The metrics used in this study were inspired by the work of Poesia et al. \cite{poesia2022synchromesh}, who generated reliable SQL queries from natural language specifications. The similarity between their research and ours prompted us to adopt their metrics. In this study, constraints that conform to the syntax and grammar of the Object Constraint Language (OCL) specifications \cite{OCLspecs} are referred to as valid constraints. OCL constraints that have been successfully compiled by an OCL compiler are reported by the validity metric. To evaluate the syntactic validity of the generated OCL constraint, we used the UML-based Specification Environment (USE) tool \cite{RichtersUSEtool}.

In addition to syntax validation, OCL constraints must also satisfy the semantic rules provided in the model specifications. OCL constraints that have been confirmed, through manual verification, to adhere to the given specification are referred to as semantically correct constraints and reported by the execution accuracy metric. Assuming that the ground-truth data is accurate and the expert judgement is sound and reliable, the manual analysis begins by matching between the generated OCL constraint and the ground truth. If the results differ, and in the case of missing ground truth, we rely on expert judgement, more specifically, the validators in this study. 

The judgement is carried out by two software engineering students: a master’s student and a Ph.D. student. Both students are conducting research involving domain modeling and constraint validation. They have undergone a thorough test of their skills in writing OCL constraints. During the analysis, each student independently verifies the OCL constraints and reports their scores. We measured the inter-rater agreement score using Cohen’s kappa coefficient \cite{cohen1960coefficient}.

\section{Empirical Evaluation}

\begin{figure}
    \centering
    \includegraphics[width=\columnwidth]{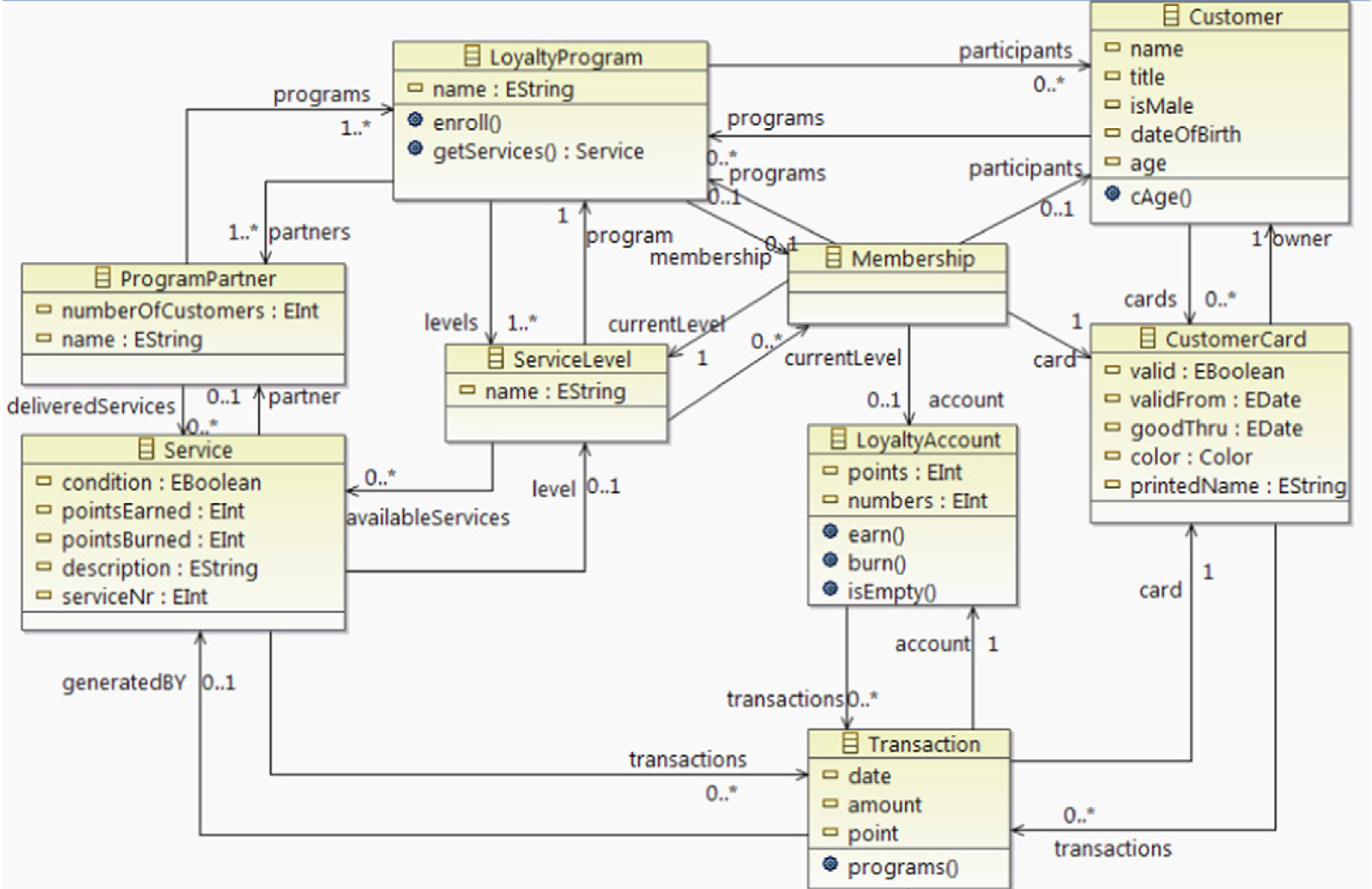}
    \caption{The Royal \& Loyal classes and associations.}
    \label{figure:royalandloyal}
\end{figure}

This section answers the proposed research questions by discussing the OCL constraints produced by Codex. The experiments were carried out using the Codex engine “code-davinci-002” with the default configurations. In order to eliminate randomness and to ensure consistency in the generated OCL constraints during the replication of the study, we set the temperature parameter to zero. We selected the double-slash character as the restart and stop sequence, as described in Section III. Since the ground-truth OCL constraints in the dataset are relatively small, we decided to keep the token limit at its default value of 256. 
\newline
\newline
\textbf{The Royal \& Loyal Model}. We present the Royal \& Loyal model as our case study, given its common use in the literature \cite{Kleppe_Warmer, bajwa_article, Wahler2008UsingPT}. The Royal \& Loyal (R\&L) model is a hypothetical company that manages loyalty programs for other businesses \cite{Kleppe_Warmer}. Figure \ref{figure:royalandloyal} illustrates the class diagrams and associations of the model \cite{bajwa_article}. We included 26 OCL constraints that were previously reported by \cite{bajwathesis, Wahler2008UsingPT} with their natural language specifications in our dataset. We will discuss the results by providing examples from the case study throughout the experiments for further insights. An example of the model is presented in Listing \ref{listing:prompt-uml-zero}.

\subsection{\textbf{RQ1. What is the impact of UML-enriched prompts on the generated OCL constraints?}}
To study the impact, we started by evaluating the results of basic prompts, and after that, we proceeded with the results of UML-enriched prompts. The syntactic validity metric was reported by compiling the generated OCL constraints in the USE modelling tool \cite{GOGOLLA200727}. The validity metric was measured by calculating the proportion of successfully compiled OCL constraints over the total number of specifications in the dataset. The execution accuracy metric was reported by manually evaluating the generated OCL constraints with a group of two students. We measured the execution accuracy metric by calculating the proportion of OCL constraints that satisfy their specifications over the total number of syntactically valid constraints. 


\subsubsection{\textbf{Basic Prompts}}
In our initial experiment, we assessed the reliability of OCL constraints generated by Codex without access to information about the UML classes and associations of the models.

The total number of specifications used in the experiment is 168. However, a proportion of the responses returned by Codex were empty OCL constraints, with 33 missing constraints observed, representing 19.5\% of the total responses. It is possible that the empty responses may have been caused by a disruption in the API connection, unstable performance from the model, or the lack of sufficient context.

We compiled the results using the USE tool and measured the validity metric, including the missing constraints, to maintain accurate measurements and a fair comparison with subsequent experiments. Then, each student manually evaluated the valid OCL constraints independently, and both reached the same conclusion, resulting in a perfect agreement score of Cohen's kappa (k = 1) \cite{cohen1960coefficient}. The total execution accuracy score achieved was 83.3\%. Table \ref{table:prompts-statistics} summarizes the results of the basic prompts. The scores suggest that Codex may have been fine-tuned on the models during the training phase, as some of them can be found on GitHub repositories.

After analyzing the experiment results more closely, we found several factors that contributed to the low validity score. The main reason was the “undefined operations”. This error occurs when there is a mismatch in referencing an attribute, operation, or association from the model in the OCL expression. This type of error accounted for 49.1\% of all observed errors. Another error category is related to the context of the generated OCL constraint. We found OCL constraints that were generated without a model context, which we referred to as “No Context” errors, and some OCL constraints were generated with incorrect referencing to the model's name, which we referred to as “Incorrect Context” errors. The top-3 syntax errors reported by the USE modeling tool are shown in Table \ref{table:basic-error-types}. 

Based on these findings, it appears that Codex may struggle with generating OCL constraints without access to UML information. The majority of the errors were either due to a mismatch in referencing or related to the context of the model. Listing \ref{listing:attribute-mistake-nouml} shows an example of this issue from the Royal \& Loyal case study, where the attribute “points” of the Transaction class was mistakenly used instead of “point”.



\begin{lstlisting}[caption={Undefined operation in referencing the attribute “points” instead of “point”.}, label={listing:attribute-mistake-nouml}, captionpos=b, float=b]
-- Specification

There must be at least one transaction for a customer card with at least 100 points.

-- Undefined operation named 'points' in expression Transaction.points()'.

context CustomerCard 
  inv:
    self.transactions->exists(t | t.points >= 100)
\end{lstlisting}



\begin{table}[t]
\centering
\caption{Basic prompt: Top-3 syntax errors. The total number of incorrect constraints is 150.}
\label{table:basic-error-types}
\resizebox{.82\columnwidth}{!}{%
\begin{tabular}{@{}|lll|@{}}
\toprule
\multicolumn{3}{|c|}{\textit{\textbf{Basic Prompt}}}                                                                                                  \\ \bottomrule \midrule
\multicolumn{1}{|l|}{\textit{\textbf{Undefined Operation}}} & \multicolumn{1}{l|}{\textit{\textbf{No Context}}} & \textit{\textbf{Incorrect Context}} \\ \midrule
\multicolumn{1}{|l|}{49.1\%}                                & \multicolumn{1}{l|}{22.8\%}                       & 11\%                                \\ \bottomrule
\end{tabular}%
}
\end{table}


When it comes to execution accuracy, the semantic errors were often due to an interpretation of the given specification. This suggests that Codex may face difficulty in fully capturing the intent without access to the model information. Therefore, it is crucial to include UML information in the prompt design to generate precise and comprehensive OCL constraints matching the specification.

It is important to note that all of the correct OCL constraints generated in this experiment were invariants, and no correct pre- or post-conditions were generated. This may be because pre- and post-conditions are behavioural constraints that often require specific knowledge about the model. This information may be more difficult for Codex to deduce without access to the UML information of each model.


\subsubsection{\textbf{UML Information \& Zero-Shot Prompts}}

In the second experiment, we assessed the impact of incorporating UML information into the prompt on the validity and execution accuracy of the generated OCL constraints. To accomplish this, we formulated the prompts to include the UML classes and associations of the model, represented in the syntax of PlantUML \cite{plantuml}.

Our analysis indicated that enriching the prompt with UML information significantly improved the validity score compared to the first experiment, from a total of 10.6\% up to 48.5\%. However, the execution accuracy of the valid constraints was slightly lower than in the first experiment, at a total of 73.1\%. These results are found in the middle of Table \ref{table:prompts-statistics}. It is possible that this is because a higher proportion of constraints were discovered to contain semantic errors as the number of syntactically valid constraints increased. The “undefined operation” error from the previous experiment as shown in Listing \ref{listing:attribute-mistake-nouml} was generated correctly in this experiment, as can be seen in Listing \ref{listing:correct-attrb-umlzero}.


\begin{lstlisting}[float=b, caption={Correctly referencing the attribute “point” with the UML-enriched and zero-shot prompts.}, label={listing:correct-attrb-umlzero}]
-- Specification

There must be at least one transaction for a customer card with at least 100 points.

-- Generated OCL 

context CustomerCard 
  inv:
    transactions->size() > 0 and transactions->exists(t | t.point >= 100)
\end{lstlisting}


Upon further analysis of the results, it was found that incorporating UML information into the prompts had a positive impact on the validity of the generated OCL constraints. Specifically, all OCL constraints in this experiment were generated within their intended context, which is in contrast to the first experiment, where a significant percentage of the constraints were generated without context. The number of empty responses was reduced from 33 (19.5\%) in the first experiment to 5 (2.9\%) in this experiment. In addition to the previously mentioned reasons in the first experiment, providing the context as UML may have also contributed to the reduction in generating empty responses. As a result, more valid constraints were generated.

Furthermore, we observed changes in the top three errors. A new error type, we refer to it as “IterExp Incorrect Source”, was identified. This error indicates that the input source to an iterator, such as “forAll()”, is not a collection. The percentage of incorrect context significantly decreased, while “undefined operation” remained the top error with the higher percentage. Table \ref{table:umlzero-error-statistics} displays the top three syntax errors and their corresponding percentages.


\begin{table}[t]
\centering
\caption{UML-enriched prompt with zero-shot learning: Top-3 syntax errors. The total number of incorrect constraints is 86.}
\label{table:umlzero-error-statistics}
\resizebox{\columnwidth}{!}{%
\begin{tabular}{@{}|lll|@{}}
\toprule
\multicolumn{3}{|c|}{\textit{\textbf{UML Information \& Zero-Shot Prompt}}}                                                                                                  \\ \bottomrule \midrule
\multicolumn{1}{|l|}{\textit{\textbf{Undefined Operation}}} & \multicolumn{1}{l|}{\textit{\textbf{IterExp Incorrect Source}}} & \textit{\textbf{Incorrect Context}} \\ \midrule
\multicolumn{1}{|l|}{64.6\%}                                & \multicolumn{1}{l|}{6.8\%}                       & 2.4\%                                \\ \bottomrule
\end{tabular}%
}
\end{table}

In contrast to the results of the first experiment, we found that the second experiment produced only a few correct pre- and post-conditions. Nonetheless, the conditions that were generated were both syntactically and semantically accurate. These results were based on models other than the Royal \& Loyal case study utilized in our dataset since it lacked examples of pre- and post-conditions. This suggests that providing additional UML information in the prompts can assist Codex in understanding the context in which the OCL specification is meant to be employed, resulting in more precise pre- and post-conditions. Listing \ref{listing:umlzero-prepost} shows an instance of a correctly generated pre- and post-condition.


\begin{lstlisting}[float=b, caption={An example of correctly generated pre-and -post condition using UML-enriched prompts with zero-shots.}, label={listing:umlzero-prepost}]
-- Specification

release(...) removes quantity of the product from the stock.

-- Generated OCL

context Product::release(qty: Integer): Boolean
    pre: qty >= 0
    post: self.stock = self.stock@pre - qty
\end{lstlisting}


\subsubsection{\textbf{UML Information \& Few-Shot Prompts}}

In our last experiment, we investigated the impact of using few-shot learning on OCL constraints generated by Codex. We used the specifications and their corresponding OCL constraints as examples for each model in the prompt. The few-shots were retrieved from the dataset, and if the ground truth was missing, we used the correctly generated constraints from the previous experiments. We evaluated the influence of these examples on the validity and execution accuracy scores of the results.

The total validity score slightly increased to 53.2\%, while the execution accuracy remained around 73.3\%, as observed in the previous experiments. The findings are presented in Table \ref{table:prompts-statistics}, indicating that few-shot learning methods have the potential to improve the reliability of OCL constraints generated by Codex. Listing \ref{listing:correct-umlfew} shows an example from the Royal \& Loyal case study that was successfully generated only during this experiment. Comparing these results with those of the previous experiments, we observed that no OCL constraints were generated with empty responses. This could be related to the same reasons as previously mentioned but improved with context and examples. The top three syntax errors remained the same as in the previous results, as summarized in Table \ref{table:umlfew-error-statistics}.


\begin{lstlisting}[float=t, caption={A correctly generated OCL constraint using a UML-enriched prompt with few-shot examples.}, label={listing:correct-umlfew}]
-- Specification

The number of valid cards for every customer must be equal to the number of programs in which the customer participates.

-- Generated OCL

context Customer
  inv:
    self.cards->select(valid = true)->size() = self.programs->size()
\end{lstlisting}

\begin{table}[b]
\centering
\caption{UML-enriched prompt with few-shot learning: Top-3 syntax errors. The total number of incorrect constraints is 78.}
\label{table:umlfew-error-statistics}
\resizebox{.9\columnwidth}{!}{%
\begin{tabular}{@{}|lll|@{}}
\toprule
\multicolumn{3}{|c|}{\textit{\textbf{UML Information \& Few-Shot Prompt}}} \\ \bottomrule \midrule
\multicolumn{1}{|l|}{\textit{\textbf{Undefined Operation}}} & \multicolumn{1}{l|}{\textit{\textbf{IterExp Incorrect Source}}} & \textit{\textbf{Incorrect Context}} \\ \midrule
\multicolumn{1}{|l|}{81\%}                                & \multicolumn{1}{l|}{8.8\%}                       & 3.7\%                                \\ \bottomrule
\end{tabular}%
}
\end{table}

\begin{table}[t]
\centering
\caption{An overview of the validity and execution accuracy percentages for all experiments.}
\label{table:prompts-statistics}
\resizebox{.95\columnwidth}{!}{%
\begin{tabular}{@{}|l|l|l|@{}}
\toprule
\multicolumn{1}{|c|}{\textit{\textbf{Prompt Design}}}       & \textit{\textbf{Validity (\%)}} & \textit{\textbf{Execution Accuracy (\%)}} \\ \bottomrule \midrule
\textit{Basic Prompt}                                       &  10.6\%                               &   83.3\%                                        \\ \bottomrule \midrule
\textit{UML Info. \& Zero-Shot}                             &  48.5\%                               &   73.1\%                                        \\ \bottomrule \midrule
\multicolumn{1}{|c|}{\textit{UML Info. \& Few-Shot}} &  53.2\%                               &   73.3\%                                        \\ \bottomrule
\end{tabular}%
}
\end{table}


\subsection{\textbf{RQ2. How similar are the generated OCL constraints by Codex to the ones written by humans?}}

The objective of this research question is to assess how natural the generated OCL constraints are to the human written constraints. This is achieved by examining the degree of embedding similarity between the generated OCL constraints and the ground truth in the dataset. We use cosine similarity as the similarity metric. 


We chose to use human-written OCL constraints as the reference for our study. This decision was based on the fact that the majority of the models in our dataset come from educational resources. As a result, we assume that the OCL constraints are written in a clear and concise manner that is easily understandable to readers such as students and practitioners. As discussed in Section III, not all natural language specifications in our dataset have corresponding OCL constraints. For this reason, these cases were not included in the analysis, and a total of 114 specifications were used in this research question. 

Similar to the approach taken by Poesia et al. \cite{poesia2022synchromesh}, we utilized a Transformer-based language model, specifically the “All-MiniLM-L6-v2” mini model, to compute the embeddings of the OCL expressions. This model was chosen for its compact size and fast processing speed \cite{MiniLM}, making it suitable for computing the embeddings of both the generated OCL expressions and the ground truth. We loaded the pre-trained model provided by the sentence-transformers (S-BERT) framework \cite{reimers-2019-sentence-bert}.

\subsubsection{\textbf{Basic Prompts}}

In the first experiment, out of 18 valid generated OCL constraints, 15 were found to be semantically correct. By analyzing the OCL constraints, it was found that the overall average cosine similarity score for the dataset was 0.74. These results suggest that the majority of the generated OCL constraints are quite similar to the ground truth, indicating that they are comprehensible and easily readable, given that the ground-truth OCL constraints are used for educational purposes. Listing \ref{listing:rq2-basic-prompt} shows an example where the only difference between the generated OCL constraint and the ground truth is the “isEmpty()” statement as the antecedent part of the implication statement. The cosine similarity score is shown at the top of Table \ref{table:rq2-exp-stats}.


\begin{lstlisting}[float=b, basicstyle=\tiny, caption={A generated OCL constraint similar to the ground truth from the Royal \& Loyal case study.}, label={listing:rq2-basic-prompt}]
-- Specification

The name and relationship of dependents is a partial identifer: they are unique among all dependents of an employee.

-- Generated OCL

context Employee
    inv:
      self.dependents->forAll(d1, d2 | d1 <> d2 implies (d1.name <> d2.name or d1.relationship <> d2.relationship))

-- Ground truth OCL

context Employee 
  inv:
    self.dependents->notEmpty() implies self.dependents->forAll(e1, e2 | e1 <> e2 implies (e1.name <> e2.name or e1.relationship <> e2.relationship))

-- Cosine similarity score = 0.9154841303825378
\end{lstlisting}


\subsubsection{\textbf{UML Information \& Zero-Shot Prompts}}

As noted in the second experiment of the first research question, there was a significant increase in the total number of valid OCL constraints generated. Out of the 82 valid generated OCL constraints, 60 were found to be semantically correct. The average cosine similarity score across all correctly generated OCL constraints was 0.61, as shown in the middle of Table VII. The lower cosine similarity scores compared to the previous experiment can be attributed to the increase in the number of valid OCL constraints generated by Codex. However, this can also suggest that the generated OCL constraints are less similar to the ground truth. Listing \ref{listing:umlzero-sim-example} provides an example of a valid alternative OCL constraint that is a correct logical solution to the one written by the human. Here, the difference in word embeddings results in a cosine similarity score of 0.72.




\begin{lstlisting}[basicstyle=\tiny, float=b,caption={A different OCL implementation generated for the same specification.}, label={listing:umlzero-sim-example}]
-- Specification

A company has at least one employee older than 50.

-- Generated OCL 

context Company 
  inv:
    self.employees->exists(age > 50)

-- Ground truth OCL

context Company
  inv:
    self.employees->select(age > 50)->notEmpty()

--Cosine similarity score = 0.7281637191772461
\end{lstlisting}


\subsubsection{\textbf{UML Information \& Few-Shot Prompts}}

In our last experiment, Codex produced 90 valid OCL constraints. Out of those, 66 were semantically accurate, slightly higher than the previous experiment. The cosine similarity score increased overall, reaching an average score of 0.80, as seen at the bottom of Table \ref{table:rq2-exp-stats}.

This result indicates that the similarity between the generated OCL constraints and the human-written ones increases by including UML information in the prompts and using few-shot examples. The cosine similarity scores are indicative of this advantageous effect.

\begin{figure}
    \centering
    \includegraphics[width=\columnwidth]{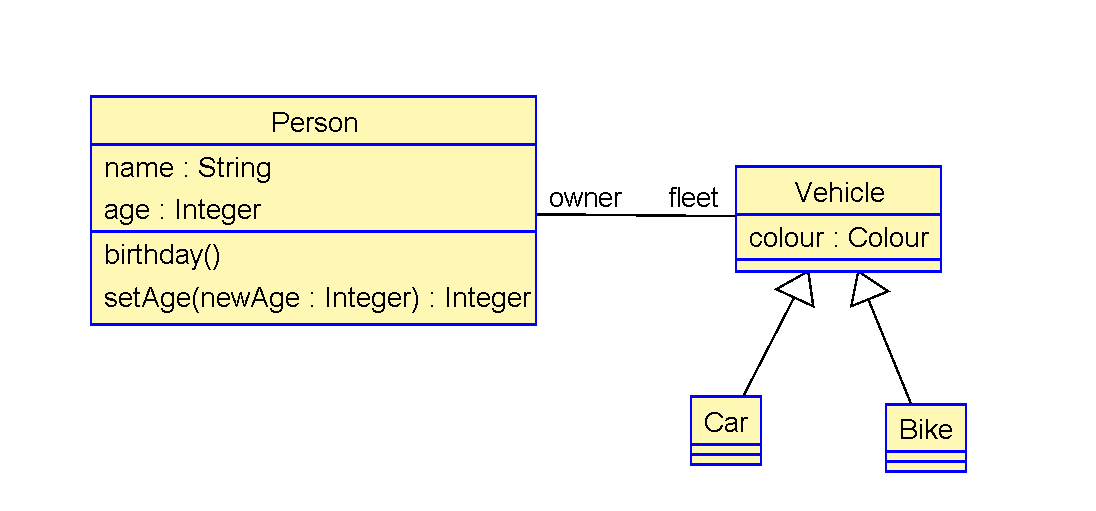}
    \caption{The UML class and associations of the Vehicle model.}
    \label{figure:vehicle-model}
\end{figure}

Listing \ref{listing:umlfew-sim} shows an example of a generated OCL constraint that uses the “select()” iterator instead of “forAll()” to apply the specification. The generated OCL constraint also appropriately uses “oclIsType()” based on the inheritance relationship in the UML diagram of the model, as shown in Figure \ref{figure:vehicle-model}.


\begin{lstlisting}[basicstyle=\tiny, float=b, caption={UML-enriched prompt with different implementation using few-shot settings.}, label={listing:umlfew-sim}]
-- Specification

A person younger than 18 owns no cars.

-- Generated OCL

context Person
  inv:
    self.age < 18 implies self.fleet->select(c | c.oclIsType0f(Car))->size() = 0

-- Ground truth OCL 

context Person
  inv:
    self.age < 18 implies self.fleet->forAll(v | not v.oclIsKindOf(Car))

-- Cosine similarity score = 0.7955201864242554
\end{lstlisting}

\begin{table}[t]
\centering
\caption{Summary of the similarity scores for all experiments.}
\label{table:rq2-exp-stats}
\resizebox{0.85\columnwidth}{!}{%
\begin{tabular}{@{}|l|l|@{}}
\toprule
\multicolumn{1}{|c|}{\textit{\textbf{Prompt Design}}}       & \textit{\textbf{Cosine Similarity Score}} \\ \bottomrule \midrule
\textit{Basic Prompt}                                       &  0.74                                 \\ \bottomrule \midrule
\textit{UML Info. \& Zero-Shot}                             &  0.61                                      \\ \bottomrule \midrule
\multicolumn{1}{|c|}{\textit{UML Info. \& Few-Shot}} &  0.80        \\ \bottomrule
\end{tabular}%
}
\end{table}


%

\section{Discussion}
This study provides statistical insights into the OCL constraints generated by Codex. Specifically, we evaluate the reliability and naturalness (with respect to human written constraints) of the generated OCL constraints using different prompt designs, ranging from basic prompts with no UML information and examples to UML-enriched prompts with zero- and few-shot examples.

Our analysis of the results from each experiment shows that basic prompts result in a low validity score for the generated OCL constraints but relatively high execution accuracy. However, syntax errors are often related to information found in the UML of the models, suggesting that the lack of this information has a negative impact on the results. The percentage of correct constraints generated by Codex suggests that the model may have already been trained on some of the models during its training phase, possibly due to the presence of these models on GitHub repositories.

In contrast, enriching the prompts with UML information significantly increases the validity score of the generated OCL constraints and results in high execution accuracy for both zero- and few-shot learning methods. This finding suggests that including information and examples regarding the models in the prompts guides Codex in generating correct OCL constraints, as also observed by the decrease in the percentage of the top syntax errors. In addition to the top three errors listed in the tables, we identified other errors reported by the tool. These include mismatches between the expected and given data types, such as assigning a string value to an integer. We also found errors that occur due to missing the “else” keyword in if-statements.

Our analysis of the OCL constraints generated by Codex shows that they are comparable and similar to the human-written constraints. This indicates that Codex has the capacity to produce human-readable OCL constraints. These results imply that Codex could be a valuable resource for model developers, as it could alleviate some of the challenges of writing OCL constraints, especially for complex models. Furthermore, the ability of Codex to produce OCL constraints from natural language specifications without necessitating substantial engineering or model transformation techniques makes it a promising black-box tool. This feature could contribute to the wider adoption of OCL as a constraint language in software modeling.


In the future, we intend to examine various prompting strategies to generate more reliable OCL constraints. However, as shown in this work, relying solely on the results of Codex is insufficient and limited. Accordingly, it is necessary to further explore automatic program repair (APR) techniques to establish a robust framework for generating OCL constraints from natural language specifications. Another direction would be to design test suites that automatically evaluate the semantics of the generated OCL constraints. This could be addressed by exploring the use of model checkers and constraint solvers.  A comprehensive comparison of different approaches to measuring the similarity between the generated OCL constraints and the ground truth can be conducted to gain a better understanding of the naturalness aspect of the results when compared to human-written constraints.



\section{Related Works}
There have been efforts to develop automated approaches for translating natural language specifications into Object Constraint Language (OCL) constraints. Unfortunately, the tools from previous works were unavailable, which prevented us from including their techniques as baselines in our evaluation. As a result, we were unable to compare our results with theirs.

\textbf{Automatic OCL generation.} Wahler proposed COPACABANA \cite{Wahler2008UsingPT}, a pattern-based and semi-automatic tool that supports model developers in identifying and specifying missing constraints in class models. This is achieved through the use of graphical models and constraint patterns. The method consists of four phases: constraint elicitation, constraint specification, consistency analysis, and code generation and tool support. Although this approach enables the identification of missing constraints, it requires partial automation and human intervention during the translation process to extract information from specifications. Furthermore, it has limitations in supporting several OCL elements, such as the iterator “select()” and the tuple datatype.

Similarly, Bajwa et al. proposed NL2OCLviaSBVR \cite{bajwa_article}, an MDA approach for automatically generating OCL constraints from natural language specifications. This approach pre-processes the input through lexical, syntax, and semantic analysis to generate a high logic representation according to the Semantics of Business Vocabulary and Business Rules (SBVR) specifications. The SBVR representation can then be transformed into another formal language, such as OCL, through the use of model transformation techniques. Although this approach is fully automated compared to Wahler \cite{Wahler2008UsingPT}, it introduces an additional layer of transformation from natural language to SBVR and then from SBVR to OCL, using the transformation library Simple Transformation (SiTra). This can introduce ambiguity when deciding which SBVR maps to the target OCL expression. Additionally, it does not support some OCL elements such as “collect()” as well as the tuple datatype.

Salemi et al. proposed En2OCL \cite{En2OCL}, an enhancement on the previous methods by addressing the limitations found in the mapping rules used during the transformation technique in NL2OCLviaSBVR \cite{bajwa_article}. This is done specifically by incorporating multiple mapping rules to the target output, and the model transformation language used is Atlas Transformation Language (ATL). However, it still imposes the same limitations in supporting tuple data types.

\textbf{Code generation.} Large language models have shown impressive findings in generating code for other programming languages. In the study by Chen et al. \cite{chen2021evaluating}, the authors fine-tuned Codex, a GPT-3 descendant model, on a corpus of publicly available code from GitHub. The authors evaluated Codex in generating code, specifically for Python programming language, and compared its performance to its ascendant, GPT-3. Their results showed that Codex excels on code generation tasks, making it a suitable candidate to explore other programming languages.

Nhan et al. \cite{nguyen2022empirical} conducted an empirical study to evaluate the quality of code suggestions generated by the AI-pair programmer Copilot, which runs a distinct version of Codex. They evaluated the correctness and understandability of the code suggestions on 33 LeetCode questions, covering various programming languages such as Java, C++, and JavaScript. Their results were consistent with Chen et al \cite{chen2021evaluating} regarding Python programming language. The authors reported that Java suggestions have the highest correctness score, and JavaScript has the lowest among the programming languages. However, they also suggested that Copilot has shortcomings, such as in the complexity of the suggested solutions and their reliance on undefined methods.

\section{Threats to Validity}
In this section, we discuss potential threats that may impact the validity of our empirical study.

\textbf{Experiments.} In some cases, empty responses were returned in our results. We conducted the experiments in a single run without performing individual queries for these empty responses.

\textbf{Quality of the dataset.} The dataset used in this study primarily consists of educational examples and may not reflect the complexity of real-world systems. As a result, it is possible that the dataset may contain missing constraints and imperfections in the implementation of the UML models, natural language specifications, and OCL constraints.

\textbf{Representation of UML models.} The UML information in the prompts was manually represented in the PlantUML language, which introduces the possibility of human error in the creation of the UML diagrams. This may result in inaccuracies or inconsistencies in the information provided by the model in the prompts.

\textbf{USE modelling tool.} The USE modelling tool, as mentioned on the GitHub repository, is a research prototype \cite{GOGOLLA200727}. As such, it may not be a fully developed, tested, or stabilized product. This presents the potential for unexpected behaviors or limitations that might affect the results of the study.

\textbf{Evaluation of execution accuracy.} Manually evaluating the semantics of the generated OCL constraints introduces the possibility of human error and bias in the evaluation process. This manual evaluation process may lead to inaccurate or inconsistent results.

\section{Conclusion}
In conclusion, this empirical study presents statistical insights into the reliability of the OCL constraints generated by Codex and their similarity to the human-written constraints. We gathered a dataset of 15 models with 168 OCL specifications to assess the impact of using different prompt designs, ranging from basic prompts with no UML information and examples to UML-enriched prompts in both zero- and few-shot learning settings. Our findings suggest that basic prompts are insufficient to generate reliable OCL constraints. However, when the prompts are enriched with the UML information of the models, the validity score of the generated OCL constraints significantly increases, resulting in high execution accuracy for both zero- and few-shot learning methods.

We also examined the naturalness of the correctly generated OCL constraints to the constraints written by humans. Our observations indicate that, on average, the generated OCL constraints are similar to their ground truth in the dataset. We assume that the human-written references are clear and easy to read, as they are collected from educational resources.

Based on our findings, we suggest that relying solely on Codex has its limitations. Therefore, we propose that incorporating automatic program repair techniques could facilitate the integration and adoption of Codex as a tool to assist model developers in writing OCL constraints to increase their productivity.

\section*{Acknowledgment}
    This work is partially supported by the Fonds de Recherche du Québec (FRQ), the Canadian Institute for Advanced Research (CIFAR), and the National Science and Engineering Research Council of Canada (NSERC).

\balance
\bibliography{bibliography}
\bibliographystyle{IEEE/IEEEtran.bst}

\end{document}